\documentclass{amsart}

\input{epsf}

\theoremstyle{definition}

\theoremstyle{remark}

\numberwithin{equation}{section}

\begin{document}

\title[Scalar Field with Robin Boundary Conditions ...]
{Scalar Field with Robin Boundary Conditions in the Worldline Formalism}%
\author
{Fiorenzo Bastianelli${}^\dagger$, Olindo Corradini${}^\dagger$ 
and Pablo A.G.\ Pisani${}^\S$}

\address{${}^\dagger$ Dipartimento di Fisica, Universit\`a di Bologna and INFN, Sezione di
Bologna; Via Irnerio, 46 - Bologna I-40126, Italy.}
\address{ ${}^\S$
  Theoretisch-Physikalisches 
Institut, Friedrich-Schiller-Universit\"at Jena; Max-Wien-Platz 1, 07743 Jena,
Germany.
Permanent address at IFLP (CONICET) and Departamento de F{\'\i}sica,
Facultad de Ciencias Exactas, Universidad Nacional de La Plata; C.C. 67 - La
Plata 1900, Argentina}



\begin{abstract}

The worldline formalism has been widely used to compute physical quantities
in quantum field theory. However, applications of this formalism to
quantum fields in the presence of boundaries have been studied only
recently.  In
this article we show how to compute in the worldline approach the heat kernel
expansion for a scalar field with boundary conditions of Robin type. In order
to describe how this mechanism works, we compute the contributions due to the
boundary conditions to the coefficients $A_1$, $A_{3/2}$ and $A_2$ of the heat
kernel expansion of a scalar field on the positive real line.

\end{abstract}

\maketitle
\section{Introduction}

Many problems in quantum field theory can be studied in the worldline
formalism. This approach provides an intuitive and, in many cases, more
efficient method to carry out computations in the 1-loop approximation
(see \cite{Fliegner:1994zc,Schubert:2001he,Bastianelli:2006rx} and
references therein.)

\smallskip

This formalism makes use of the connection between some quantities in
quantum field theory, like the effective action, and the operator $e^{- T
A}$ corresponding to some relevant differential operator $A$. The first
important step in this approach is to identify a classical particle
whose Hamiltonian, after first quantization, is this differential operator
$A$. If this identification is found, then the heat kernel corresponds to
the evolution operator of the particle in Euclidean time $ T$. The kernel
$\langle y|e^{- T A}|x\rangle$ of this operator $e^{- T A}$ is a quantum
mechanical transition amplitude and therefore can be also computed in the path
integral formulation of first quantization.

\smallskip

It is important to mention that many physical quantities of the quantum field
theory are determined by the small-$ T$ asymptotic expansion of the
heat-kernel $\langle y|e^{- T A}|x\rangle$. In this sense, the path integral
computation of this transition amplitude results particularly appropriate because the
small-$ T$ expansion comes out in a natural way. By this procedure
the worldline formalism has been proved useful in the study of quantum fields
of different spins and their coupling to external gravity.

\smallskip

The purpose of this article is to describe how this formalism can also be
applied to quantum field theory on manifolds with boundaries. In order to
do this, one has to deal with the path integral formulation of the
quantum mechanics of a point particle on a manifold with boundaries. The
main difficulty is to implement the boundary conditions imposed on the
quantum fields, i.e. on the domain of the differential operator $A$, in the
path integral calculation of the quantum mechanical transition amplitude
$\langle y|e^{- T A}|x\rangle$. This problem has been solved
in \cite{Bastianelli:2006hq} for Dirichlet and Neumann boundary conditions by
the method of images. In this article we show how to implement Robin
boundary conditions in the path integral calculation and we compute the
contributions of the boundary conditions to the coefficients $A_1$, $A_{3/2}$
and $A_2$ of the small-$ T$ asymptotic expansion of the heat kernel for
a scalar field on the positive half-line. For this setting, Robin
boundary conditions are the most general boundary conditions determined by
the selfadjointness of the differential operator $A$.

\smallskip

The method of images is not straightforwardly applicable to Robin boundary
conditions. However, as it is shown in the appendix (see
also \cite{Clark:1980xt}), one can implement Robin boundary conditions on
the heat kernel by using the image construction for Neumann boundary
conditions supplemented by a suitable delta-function potential at the
boundary. This delta-function then constrains the paths which contribute to
the path integral computation of the transition amplitude.

\smallskip

In the next section, we describe the worldline method by considering
the heat-kernel trace for a scalar field on a manifold without boundaries.
In section \ref{Robin} we apply the worldline approach to a scalar field on
the positive half-line under Robin boundary conditions at the origin and
we compute the contributions of the boundary to the first coefficients of the
heat kernel expansion. In section \ref{conclusions} we draw our conclusions and
comment on possible generalizations of our technique.

\section{Heat-kernel expansion without boundaries}

In order to describe the worldline approach to quantum field theories let us
consider the 1-loop effective action of a scalar field $\phi(x)$ with
a self-interaction $U(\phi(x))$ on a flat manifold $\mathcal{M}$. The classical
action $I[\phi(x)]$ of the scalar field is given by
\begin{equation}
	I[\phi(x)]=\int_{\mathcal{M}}dx\,\left\{\frac{1}{2}
(\partial\phi(x))^2+U(\phi(x))\right\}
\end{equation}
and the 1-loop effective action can be represented as
\begin{eqnarray}\label{effact}
	\Gamma[\phi(x)]&=&I[\phi(x)]
	-\hbar\log{\rm Det}^{-1/2}\left\{-\Delta+U''(\phi(x))\right\}=
	\nonumber\\
	&=&I[\phi(x)]-\frac{\hbar}{2}\int_0^\infty \frac{d T}{ T}\,
	{\rm Tr}\left(e^{- T
	\left\{-\Delta+U''(\phi(x))\right\}}\right)\nonumber=\\
	&=&I[\phi(x)]-\frac{\hbar}{2}\int_0^\infty \frac{d T}{ T}\,	
	\int_{\mathcal{M}}dx\,\langle x|e^{- T
	\left\{-\Delta+U''(\phi(x))\right\}}|x\rangle
\end{eqnarray}
The integrand in the last expression is the heat kernel
\begin{equation}\label{heaker}
	\langle y|e^{- T\left\{-\Delta+U''(\phi(x))\right\}}|x\rangle
\end{equation}
of the Schr\"odinger operator
\begin{equation}\label{ham}
	-\Delta +U''(\phi(x))
\end{equation}
evaluated at the diagonal $x=y$. Inspection of expression (\ref{effact}) shows
that the small-$ T$ asymptotic expansion of this heat-kernel
contains information about the ultraviolet divergences of the theory.

\smallskip

In the worldline approach, the kernel (\ref{heaker}) is regarded as the
quantum mechanical transition amplitude in Euclidean time $ T$ between $x$
and $y$ of a point particle whose Hamiltonian is given by (\ref{ham}). In
fact, the Schr\"odinger operator (\ref{ham}) can be regarded as the Hamiltonian,
after first quantization, of a point particle whose Euclidean classical action
is
\begin{equation}
	S[x(t)]=\int_0^ T dt\,\left\{\frac{1}{4}\dot{x}^2(t)+V(x(t))\right\}
\end{equation}
where
\begin{equation}
	V(x):=U''(\phi(x))
\end{equation}
Therefore, we can represent the transition amplitude (\ref{heaker}) in the path
integral approach
\begin{equation}\label{patint}
	\langle y|e^{- T\left\{-\Delta +V(x)\right\}}|x\rangle =
	\int_{x(0)=x}^{x( T)=y}\mathcal{D}x(t)\,
	e^{-\int_0^ T dt\,\left\{\frac{1}{4}\dot{x}^2+V(x(t))\right\}}
\end{equation}
With this procedure we can compute some quantities in quantum field theory
in terms of quantum-mechanical path integrals. An essential step in
this description is the identification of the differential operator appearing in
the first line of expression (\ref{effact}) with the quantized Hamiltonian of a
point particle.

\smallskip

Next, we show how the path integral in expression (\ref{patint}) turns out to be
convenient for the computation of the small-$ T$ asymptotic expansion of the
transition amplitude. In order to do that, we rescale the Euclidean time
variable $t$ by introducing $\tau:=t/ T$ and we replace the integration
over trajectories $x(t)$ by an integration over the ``quantum fluctuations''
$y(\tau):=x( T\tau)-(y-x)\tau-x$
\begin{eqnarray}\label{masequ}
	\langle y|e^{- T\left\{-\Delta +V(x)\right\}}|x\rangle&=&
	e^{-\frac{(y-x)^2}{4 T}}
	\int_{y(0)=0}^{y(1)=0}\mathcal{D}y(\tau)\,
	e^{-\frac{1}{4 T}\int_0^1 d\tau\,\dot{y}^2(\tau)}\times\nonumber\\
	&&\times\ e^{- T\int_0^1 d\tau\,V((y-x)\tau+x+y(\tau))}\nonumber\\
	&=:&e^{-\frac{(y-x)^2}{4 T}}
	\left\langle e^{- T\int_0^1
	d\tau\,V((y-x)\tau+x+y(\tau))}\right\rangle\nonumber\\
	&=&e^{-\frac{(y-x)^2}{4 T}}\,\sum_{n=0}^\infty \frac{(- T)^n}{n!}
	\int_0^1 d\tau_1\ldots\int_0^1 d\tau_n\,\times\nonumber\\
	&&\times\ \left\langle\prod_{k=1}^n
	V((y-x)\tau_k+x+y(\tau_k))\right\rangle
\end{eqnarray}
Notice that we end up with a perturbative calculation in (0+1)-field theory
where the number of loops is related to the power of $ T$ and the propagator
is proportional to $ T$.

\smallskip

For our purposes it suffices to consider the trace of the heat kernel
\begin{eqnarray}\label{traheaker}
	{\rm Tr}\,e^{- T \left\{-\Delta +V(x)\right\}}
	&=&
	\int_{\mathcal M}dx\,
	\left\langle
	e^{- T\int_0^1 d\tau\,V(x+y(\tau))}
	\right\rangle
	\nonumber\\
	&=&\sum_{n=0}^\infty \frac{(- T)^n}{n!}\int_{\mathcal M}
	 dx\,
	\int_0^1 d\tau_1\ldots\int_0^1 d\tau_n\,
	\left\langle\prod_{k=1}^n
	V(x+y(\tau_k))\right\rangle
\end{eqnarray}
If the potential is an analytic function, we can make a Taylor expansion and
write the integrand in the previous expression as a sum whose terms consist, in
general, of a product between the potential and its derivatives evaluated
at $x$ and many-point functions, which, due to Wick's theorem, can be written
in terms of two-point functions $\langle y(\tau_{k_1})y(\tau_{k_2})\rangle$.
Since these propagators are proportional to $ T$ we obtain an asymptotic
expansion in integer powers of $ T$ whose coefficients are integrals on
${\mathcal M}$ of products of the potential and its derivatives.

\smallskip

Finally, we normalize the expectation value $\langle 1\rangle$ such that for the
case $V(x)\equiv 0$ we obtain
\begin{equation}\label{fre}
	\langle y|e^{- T\left\{-\Delta\right\}}|x\rangle=
	e^{-\frac{(y-x)^2}{4 T}}\cdot\langle 1\rangle=
	\frac{e^{-\frac{(y-x)^2}{4 T}}}{\sqrt{4\pi  T}}
\end{equation}
which is the known value for the transition amplitude for a free particle.
Notice that the numerator in the last expression corresponds to the contribution
of the classical trajectory whereas the denominator corresponds to
the contributions of the quantum fluctuations.

\smallskip

We end this section by recalling that for a Schr\"odinger differential operator
on a manifold with boundaries the trace of the heat kernel admits
the following small-$ T$ asymptotic expansion\footnote{This expansion holds
under certain regularity assumptions on the manifold, the potential and the
(local) boundary conditions.}
\begin{equation}\label{heakerbou}
	{\rm Tr}\,e^{- T \left\{-\Delta +V(x)\right\}}
	=\frac{1}{(4\pi T)^{m/2}}\sum_{n=0}^\infty
	A_{n/2}\cdot T^{n/2}
\end{equation}
where $m$ is the dimension of the base manifold ${\mathcal M}$. The
coefficients $A_{n/2}$ for odd $n$ vanish for manifolds without boundary (see
the discussion after eq.\ (\ref{traheaker}).) In the next section it will
become clear how these half-integer powers of $ T$ appear due to the presence
of boundaries.

\section{Heat kernel expansion under Robin boundary conditions}\label{Robin}

In this section we consider the small-$ T$ asymptotic expansion of the
trace of the heat kernel corresponding to a Schr\"odinger operator on the
positive half-line ${\mathcal M}={\mathbb R}^+$. We impose Robin
boundary conditions on the functions $\psi(x)$ in the domain of the
Schr\"odinger operator
\begin{equation}\label{rob}
	\left.\partial_x\psi(x)+\gamma\cdot\psi(x)\right|_{x=0}=0
\end{equation}
where $\gamma\in\mathbb{R}$.

\smallskip

This type of boundary conditions can be implemented in the path integral
by computing the contributions of all paths that correspond to Neumann
boundary conditions and, at the same time, introducing a Dirac delta-function
in the potential (see appendix.) The computation of the contribution of all
paths that correspond to Neumann boundary conditions can be performed by the
method of images, according to which
\begin{equation}\label{neu}
	\langle x|e^{- T\left\{-\Delta+V(x)\right\}}|x\rangle_N=
	\langle
	x|e^{- T\left\{-\Delta+V_{\mathbb{R}}(x)\right\}}
	|x\rangle_{\mathbb{R}}+\langle
	-x|e^{- T\left\{-\Delta+V_{\mathbb{R}}(x)\right\}}
	|x\rangle_{\mathbb{R}}
\end{equation}
where the subscript $N$ stands for ``Neumann'' and the subscript $\mathbb{R}$
indicates that the corresponding transition amplitudes are computed in the whole
real line. Accordingly, $V_{\mathbb{R}}(x)$ is the extension to the whole real
line of the potential $V(x)$, which is defined only on $\mathbb{R}^+$, by a
reflection about the origin. As just mentioned, in order to obtain the
transition amplitude for Robin boundary conditions we still have to include
in the potential a Dirac delta-function. Finally, using eq.\ (\ref{masequ})
to compute the transition amplitudes in the whole real line, the trace of the
heat kernel for Robin boundary conditions can be written as
\begin{eqnarray}\label{trarob}
	{\rm Tr}_\gamma\,e^{- T \left\{-\Delta +V(x)\right\}}
	&=&
	\int_0^\infty dx\,
	\langle
	x|e^{- T\left\{-\Delta+{V^\gamma}_{\mathbb{R}}(x)\right\}}
	|x
	\rangle_{\mathbb{R}}+
	\nonumber\\
	&&+\int_0^\infty dx\,
	\langle
	-x|e^{- T\left\{-\Delta+{V^\gamma}_{\mathbb{R}}(x)\right\}}
	|x\rangle_{\mathbb{R}}
	\nonumber\\
	&=&
	\sum_{n=0}^\infty \frac{(- T)^n}{n!}
	\int_0^\infty dx\,
	\int_0^1 d\tau_1\ldots\int_0^1 d\tau_n\times\nonumber\\
	&&\left\{\left\langle\prod_{k=1}^n
	{V^\gamma}_{\mathbb{R}}(x+y(\tau_k))\right\rangle
	\right.\nonumber\\ &&\left.
	+e^{-\frac{x^2}{ T}}\,
	\left\langle\prod_{k=1}^n
	{V^\gamma}_{\mathbb{R}}((1-2\tau_k)x+y(\tau_k))\right\rangle
	\right\}
\end{eqnarray}
where
\begin{equation}\label{vtil}
	{V^\gamma}_{\mathbb{R}}(x)=V(x)+\theta(-x)(V(-x)-V(x))-
	2\gamma\cdot\delta(x)
\end{equation}
The term with the Heaviside function $\theta(x)$ represents the extension of
the potential to the negative half-line whereas the term with the
Dirac delta-function is needed in order to obtain the transition function for
the boundary conditions given by (\ref{rob}). Notice that the factor
$e^{-x^2/ T}$ in the second term of eq.\ (\ref{trarob}) is responsible for
the appearance of half-integer powers of $ T$.

\smallskip

Next, we compute the contributions to the first coefficients $A_{n/2}$
of the heat kernel expansion (\ref{heakerbou}) due to Robin boundary
conditions. Since $\gamma=0$ corresponds to Neumann boundary conditions
these contributions, which are proportional to positive powers of $\gamma$,
must be added to the coefficients $A_{n/2}$ corresponding to Neumann boundary
conditions. The coefficients $A_{n/2}$ for Neumann boundary conditions have
been already computed in the worldline formalism in \cite{Bastianelli:2006hq}.

\smallskip

It is interesting to notice that the inclusion of the Dirac delta-function
leads to Dirichlet boundary conditions for $\gamma\rightarrow -\infty$. Indeed,
by the method of images, the transition function for these boundary conditions
is given by the difference of the terms at the {\small R.H.S.\ }of (\ref{neu}).
This subtraction amounts to omit from the computation the contribution of all
those paths that hit the origin at least once. In accordance, as it can
be seen from the first line of eq.\ (\ref{traheaker}), the Dirac delta-function
in (\ref{vtil}) cancels the contributions of the paths that hit the origin.

\smallskip

To compute the contributions of the boundary condition (\ref{rob}) to the
coefficients $A_{n/2}$ (with respect to Neumann boundary conditions) we do not
use the procedure described in the previous section. Due to the singularities
introduced in the potential by the Dirac delta-function and also by making the
reflection with respect to the origin, one cannot make a Taylor expansion of
${V^\gamma}_{\mathbb{R}}$ in expression (\ref{trarob}) (see
the discussion after (\ref{traheaker}).) Nevertheless, one can deal with the
Heaviside and
the delta-function introduced in (\ref{vtil}) by appropriately constraining the
paths in the functional integration. We describe this mechanism by computing the
contributions to the coefficients $A_1$, $A_{3/2}$ and $A_2$.

\smallskip

The first non-vanishing contribution due to Robin boundary conditions comes from
the term corresponding to $n=1$ in eq.\ (\ref{trarob})
\begin{equation}\label{n=1}
	- T\int_0^\infty dx\,
	\int_0^1 d\tau_1\,
	\left\{
	\Big\langle-2\gamma\cdot\delta(x+y(\tau_1))\Big\rangle
	+e^{-\frac{x^2}{ T}}\Big\langle-2\gamma\cdot
	\delta((1-2\tau_1)x+y(\tau_1))\Big\rangle
	\right\}
\end{equation}
Let us consider the delta-function in the expectation value. For the first term,
this means that one should consider the contributions of only those
paths $y(\tau)$ begining and ending at the origin in Euclidean time $T$ such
that at $\tau=\tau_1$ reach the point $-x$. This restriction can
be implemented by computing the product of the (free) transition amplitude (see
eq.\ (\ref{fre})) from the origin to $-x$ in time $ T\tau_1$ times the (free)
transition amplitude from $-x$ back again to the origin in time $ T(1-\tau_1)$.
The same calculation can be implemented for the second term in (\ref{n=1})
but considering the intermediate point $-(1-2\tau_1)x$ instead of $-x$.
The result can be written as
\begin{eqnarray}
	2 T\gamma\int_0^\infty dx\,
	\int_0^1 d\tau\,
	\left\{
	\frac{e^{-\frac{x^2}{4 T\tau_1}}}{\sqrt{4\pi T\tau_1}}\cdot
	\frac{e^{-\frac{x^2}{4 T(1-\tau_1)}}}{\sqrt{4\pi T(1-\tau_1)}}
	\right.\nonumber\\
	\left.
	\mbox{}+e^{-\frac{x^2}{ T}}\,
	\frac{e^{-\frac{(1-2\tau_1)^2
	x^2}{4 T\tau_1}}}{\sqrt{4\pi T\tau_1}}\cdot
	\frac{e^{-\frac{(1-2\tau_1)^2
	x^2}{4 T(1-\tau_1)}}}{\sqrt{4\pi T(1-\tau_1)}}
	\right\}
	=\frac{ T}{\sqrt{4\pi T}}\cdot 2\gamma
\end{eqnarray}
where the result can be easily obtained interchanging the order of integration.
We conclude that the contribution of the boundary conditions to $A_1$ is given
by $2\gamma$.

\smallskip

Next, we proceed with the term corresponding to $n=2$ in eq.\ (\ref{trarob}).
There are two contributions to be considered, one proportional to $\gamma$ and
the other one proportional to $\gamma^2$. The contribution proportional to
$\gamma^2$ is given by
\begin{eqnarray}\label{n=2g2}
	\frac{ T^2}{2}\,4\gamma^2\int_0^\infty dx\,
	\int_0^1 d\tau_1\int_0^1 d\tau_2\,
	\left\{
	\Big\langle \delta(x+y(\tau_1))\cdot\delta(x+y(\tau_2))\Big\rangle
	\right.\nonumber\\ \left.
	+e^{-\frac{x^2}{ T}}\Big\langle \delta((1-2\tau_1)x+y(\tau_1))\cdot
	\delta((1-2\tau_2)x+y(\tau_2))\Big\rangle
	\right\}\nonumber\\
	=
	4 T^2\gamma^2\int_0^\infty dx\,
	\int_0^1 d\tau_1 \int_{\tau_1}^1 d\tau_2\,
	\left\{
	\frac{e^{-\frac{x^2}{4 T\tau_1}}}{\sqrt{4\pi T\tau_1}}
	\frac{1}{\sqrt{4\pi T(\tau_2-\tau_1)}}
	\frac{e^{-\frac{x^2}{4 T(1-\tau_2)}}}{\sqrt{4\pi T(1-\tau_2)}}
	\nonumber\right.\\ \left.+
	e^{-\frac{x^2}{ T}}\,
	\frac{e^{-\frac{(1-2\tau_1)^2x^2}{4 T\tau_1}}}
	{\sqrt{4\pi T\tau_1}}
	\frac{e^{-\frac{(\tau_2-\tau_1)x^2}{ T}}}
	{\sqrt{4\pi T(\tau_2-\tau_1)}}
	\frac{e^{-\frac{(1-2\tau_2)^2x^2}{4 T(1-\tau_2)}}}
	{\sqrt{4\pi T(1-\tau_2)}}
	\right\}
	=\frac{ T^{3/2}}{\sqrt{4\pi T}}\cdot\sqrt{\pi}\gamma^2
\end{eqnarray}
Therefore, the contribution of the boundary conditions to $A_{3/2}$ is given
by $\sqrt{\pi}\gamma^2$.

\smallskip

The leading order in $ T$ of the contribution proportional to $\gamma$
in the $n=2$ term of eq.\ (\ref{trarob}) contributes to the coefficient $A_2$
and is given by
\begin{eqnarray}
	\frac{ T^2}{2}(-2\gamma)
	\int_0^\infty dx\,\int_0^1d\tau_1 \int_0^1 d\tau_2\,
	\left\{
	\Big\langle 2V(x+y(\tau_1))\,\delta(x+y(\tau_2))\Big\rangle
	\right.\nonumber\\ \left.+e^{-\frac{x^2}{ T}}\,
	\Big\langle 2V((1-2\tau_1)x+y(\tau_1))\,
	\delta((1-2\tau_2)x+y(\tau_2))\Big\rangle\right\}
\end{eqnarray}
The leading order in $T$ of this expression is obtained by making a Taylor
expansion of the potential about the origin
\begin{eqnarray}\label{n=2g1}
	-2 T^2\gamma\,V(0)
	\int_0^\infty dx\,\int_0^1 d\tau_1\int_0^1 d\tau_2\,
	\left\{
	\frac{e^{-\frac{x^2}{4 T\tau_2}}}{\sqrt{4\pi T\tau_2}}
	\frac{e^{-\frac{x^2}{4 T(1-\tau_2)}}}{\sqrt{4\pi T(1-\tau_2)}}
	\right.\nonumber\\ \left.+
	e^{-\frac{x^2}{ T}}\,
	\frac{e^{-\frac{(1-2\tau_2)^2
	x^2}{4 T\tau_2}}}{\sqrt{4\pi T\tau_2}}
	\frac{e^{-\frac{(1-2\tau_2)^2
	x^2}{4 T(1-\tau_2)}}}{\sqrt{4\pi T(1-\tau_2)}}\right\}
	=-\frac{ T^2}{\sqrt{4\pi T}}\cdot 2\gamma\,V(0)
\end{eqnarray}
This is a contribution to the coefficient $A_2$ proportional to $\gamma$.
There is another contribution to this coefficient due to the boundary
conditions which is proportional to $\gamma^3$. We end this section by computing
this contribution, which comes from the $n=3$ term in eq.\ (\ref{trarob}) and
is given by
\begin{eqnarray}
	-\frac{ T^3}{3!}(-8\gamma^3)
	\int_0^\infty\!\!\! dx \int_0^1\!\!\!\!d\tau_1
	\int_0^1\!\!\!\!d\tau_2
	\int_0^1\!\!\!\!d\tau_3\,
	\left\{
	\Big\langle\delta(x+y(\tau_1))
	\delta(x+y(\tau_2))
	\delta(x+y(\tau_3))
	\Big\rangle\right.\nonumber\\ \left.+
	e^{-\frac{x^2}{ T}}\,
	\Big\langle\delta((1-2\tau_1)x+y(\tau_1))
	\delta((1-2\tau_2)x+y(\tau_2))
	\delta((1-2\tau_3)x+y(\tau_3))
	\Big\rangle\right\}\nonumber\\
	=
	\frac{ T^2}{\sqrt{4\pi T}}\cdot\frac{4}{3}\gamma^3
\end{eqnarray}
Taking into account this last result together with the one given in eq.\
(\ref{n=2g1}) we conclude that the total contribution to $A_2$ due to the
boundary conditions is given by $2\gamma\,V_0+4/3\,\gamma^3$. All these
contributions are the expected ones (see e.g.\
\cite{Branson:1999jz}.)

\section{Conclusions}\label{conclusions}

Worldline approaches can be successfully used to study QFT's on manifolds with
boundaries. After recalling the methods employed
in \cite{Bastianelli:2006hq}, we have considered the inclusion of Robin
boundary conditions. These conditions can be enforced by adding a
delta-function potential to the setup for Neumann boundary conditions described
in \cite{Bastianelli:2006hq}. This way we have been able to recover the
expected corrections to the heat-kernel coefficients up to, and including,
$A_2$. We plan to further simplify and extend the methods reviewed
in the present paper \cite{new?}. In addition it would be interesting to extend
them to include curved boundaries, curved spaces and fields with nontrivial
spins. 

\smallskip

\noindent {\bf Acknowledgments:} P.A.G.P.\ would like to thank the
participants and the organizers of QFEXT'07. P.A.G.P.\ also acknowledges
financial support from CONICET (PIP 6160), UNLP (11/X381), INFN and DAAD.

\appendix

\section{Appendix}

Consider the following kernel, defined on the half line
\begin{equation}\label{one}
\psi(y,x;\beta):=\langle y
|e^{-\beta H_\gamma}| x\rangle_{\mathbb R} +  
\langle -y |e^{-\beta H_\gamma}| x\rangle_{\mathbb R}
\end{equation}
with 
\begin{eqnarray}
H_\gamma = -\partial^2_y-2\gamma \delta(y)~.
\end{eqnarray}
By construction, (\ref{one}) is even under $y\to -y$. In other words
$\psi(y,x;\beta)$ can be extended to the whole line such that it is continuous
as a function of $y$. On the other hand, its first derivative is odd
\begin{eqnarray}
\partial_y\psi(y,x;\beta) = -\partial_y\psi(-y,x;\beta)
\label{four}
\end{eqnarray}
and is not defined at $y=0$ because of the presence of the delta function. In
fact, (\ref{one}) satisfies the heat equation
\begin{eqnarray}
\left(\partial_\beta- \partial^2_y
-2\gamma\delta(y)\right)\psi(y,x;\beta) =0
\end{eqnarray}
which is equivalent to the heat equation for a free particle away from the
boundary whose wave function satisfies the jump condition
\begin{eqnarray}\label{jump}
-\partial_y\psi(0^+,x;\beta)+\partial_y\psi(0^-,x;\beta)
-2\gamma\psi(0,x;\beta) =0
\end{eqnarray}
Using eqs. (\ref{four}) and (\ref{jump}) we obtain
\begin{eqnarray}
\partial_y\psi(0^+,x;\beta)
+\gamma\cdot\psi(0,x;\beta) =0
\end{eqnarray}
Therefore, (\ref{one}) is the heat kernel for a free particle on the half line
subject to Robin boundary conditions. Note that taking $\gamma\to 0$ one
correctly achieves Neumann boundary conditions.



\end{document}